\documentstyle[12pt]{article}
\begin{document}
\footnotetext[1]{
Electronic address: tiitaka@postman.riken.go.jp
}

\title{
Calculating the linear response functions 
of non-interacting electrons
by the time-dependent Schroedinger equation 
}

\author{ Toshiaki Iitaka$^*$, Shintaro Nomura, Hideki Hirayama, \\
Xinwei Zhao, Yoshinobu Aoyagi, Takuo Sugano \\
Nanoelectronics Materials Group \\
Frontier Research Program, RIKEN \\
2-1 Hirosawa, Wako, Saitama 351-01, JAPAN 
}

\date{May 2, 1997}

\maketitle

\setcounter{tocdepth}{2}

\begin{flushleft}
PACS: 02.70.-c
\end{flushleft}

\clearpage

\begin{abstract}
An O(N) algorithm is proposed for calculating  linear response functions of non-interacting electrons. This algorithm is simple and suitable to parallel- and  vector- computation. 
Since it avoids \( O(N^3) \) computational effort of matrix diagonalization, it requires only \( O(N) \)  computational efforts where \( N\) is the dimension of the statevector.
The use of  this \( O(N) \) algorithm is very effective since otherwise we have to calculate large number of eigenstates, i.e., the occupied one-electron states up to the Fermi energy and the unoccupied states with higher energy. 
The advantage of this method compared to the Chebyshev polynomial method  recently developed by Wang (L.W.~Wang, Phys. Rev. B {\bf 49}, 10154 (1994);L.W.~Wang, Phys. Rev. Lett. {\bf 73}, 1039 (1994) ) is that our method can calculate linear response functions without any storage of huge statevectors on external storage.

\end{abstract}

\section{Introduction}

Computing the linear response functions (and the density of states) of large systems with thousands of atoms by using conventional methods requires to calculate the eigenvalues and eigenvectors of  \( N \times N \) Hamiltonian matrices ($N \gg 10^{6}$) from the lowest state to the Fermi energy and beyond it.
The standard diagonalization routines are too much time-consuming in treating these problems because their computing time is proportional to \( N^3\) .
Therefore efficient numerical algorithms, such as recursive Green's function methods \cite{Recursive,Branger}, the Lanczos methods \cite{Lanczos,Dagotto94,Cordelli,Jaklic}, the Chebyshev polynomial expansion  \cite{Recipes,Kosloff,Hoffman,Silver,Roder,Wang,Voter,Goedecker,Sankey}, and conjugate gradient methods \cite{Payne,Nomura96}  have been developed and applied to various problems.

In this paper, we present an efficient method for calculating the linear response functions of large quantum system.
We give up the calculation of each exact eigenstates, instead we compute linear response functions by integrating the time-dependent Schroedinger equations for a finite period determined by the required energy resolution.
Since it avoids \( O(N^3) \) computational efforts of matrix diagonalization, it requires only \( O(N) \) computational efforts for sparse Hamiltonian matrices.
To realize this scheme, we exploit several  numerical techniques such as the Chebyshev polynomial expansion of matrix functions \cite{Recipes,Kosloff,Hoffman,Silver,Roder,Wang,Voter,Goedecker,Sankey},  random state vectors \cite{Sankey,Drabold,Skilling}, Hamiltonian matrix discretized in real space \cite{Chelikowsky,Fletcher},  the time-dependent Schroedinger equation discretized in real time \cite{Askar,Lefo,Iitaka94,Iitaka94b,Iitaka95a,Iitaka95b,Iitaka96,Feit,Kuroda,Raedt1,Raedt2,Kawarabayashi,Natori,Glutsch}.

\section{Time-dependent methods}
\label{sec:time}

In this section, we describe how we reached the conclusion that
we can calculate efficiently the linear response functions of large
quantum systems by using the time-dependent homogeneous Schroedinger equations.

\subsection{Diagonalize or not diagonalize?}

Let us compare the computational efforts of the conventional diagonalization method and the time-dependent method by counting the number of floating point multiplications as a function of matrix dimension $N$, and show that time-dependent method is more efficient when large number of eigenstates are involved. 

First, we review the relation between the eigenstate representation and the time-dependent representation of linear response functions. The linear response function $\chi_{BA}(\omega+i\eta)$ of an observable $B$ due to a {\em monochromatic} perturbation $H^{ex}=e^{-i(\omega+i\eta) t}A $ is calculated by time-dependent perturbation theory \cite{Imry},
\begin{eqnarray}
\label{eq:chi.time0}
\chi_{BA}(\omega+i\eta)&=& 
\displaystyle
(-i) \int_0^\infty dt e^{+i(\omega+i\eta)t}
\left\{
\langle E_g| e^{+iHt} B e^{-iHt} A | E_g \rangle 
-c.c.
\right\} \\
\label{eq:chi.time}
 &\approx&
2 \int_0^T dt e^{+i(\omega +i\eta)t}
{\rm Im} \left\{ \langle E_g| B e^{-iHt} A | E_g \rangle e^{+iE_gt}\right\} 
\end{eqnarray}
where $|E_g\rangle$ and $E_g$ are the groundstate of the many electron system and its energy, respectively;
$\omega$ and $\eta$  are the frequency and its resolution, respectively; $T\gg 1/\eta$ is the integration time.
We use atomic units (a.u.), and indicate complex conjugate by ``c.c.''.
In numerical calculation of (\ref{eq:chi.time}), we have to discretize it in time,
e.g.,
\begin{equation}
\label{eq:chi.time.discrete}
\chi_{BA}(\omega+i\eta)=
2  \sum_{m=0}^M \Delta t e^{+i(\omega+i\eta)m \Delta t}
{\rm Im} \left\{ \langle E_g| B e^{-iHm \Delta t} A | E_g \rangle e^{+iE_gm\Delta t} \right\}
\end{equation}
where $M=T/\Delta t$ is the number of timesteps, $T$ is the integration time in (\ref{eq:chi.time}), and $\Delta t$ is the width of timestep.
On the other hand,
we obtain the eigenstate representation 
by inserting  $I=\sum_{m=1}^{N}|E_m\rangle \langle E_m| $ into  (\ref{eq:chi.time0}), 
\begin{eqnarray}
\label{eq:chi.eigen}
\chi_{BA}(\omega+i\eta)
&=&\sum_{m=1}^N 
\frac{\langle E_g| B | E_m \rangle  \langle E_m| A | E_g \rangle}{(\omega+i\eta) -(E_m-E_g )}  \nonumber \\
&& - \sum_{m=1}^N 
\frac{\langle E_g| A | E_m \rangle  \langle E_m| B | E_g \rangle}{(\omega+i\eta) +(E_m-E_g )} 
.
\end{eqnarray}

%We also assume that the Hamiltonian matrix $H$ has eigenvalues in the
%interval $[-1,1]$. We can realize this by shifting and renormalizing
%the given Hamiltonian

Next, we show the estimated computational efforts in Table~\ref{tbl:compare}. The diagonalization method for $N \times N$ Hamiltonian matrix requires memory space of $O(N^2)$ and computational effort of $O(N^3)$. On the other hand, the time-dependent method requires memory space of $O(N^2)$ and computational effort of $O(MN^2)$ where $M$ is the number of timesteps determined by the required energy resolution (See section~\ref{subsec:homo} ).
By choosing an appropriate basis set, we can make the Hamiltonian a sparse matrix having only $O(N)$ non-zero matrix elements \cite{Chelikowsky,Iitaka94b}. As the result, the computational effort and the memory space of the time-dependent method are reduced to $O(MN)$ and $O(N)$, respectively.
Thus the time-dependent method can be  more efficient than diagonalization method in large $N$ limit.

\subsection{Newton or Schroedinger?}

Table~\ref{tbl:equations} classifies various time-dependent methods in terms of kind of equation and homogeneity.  Though the Newton equations of harmonic oscillators \cite{Williams,Yakubo,Fukamachi,Tanaka,Hukushima} are mathematically equivalent to the Schroedinger equations in the eigenstate representation, use of the Schroedinger equation \cite{Askar,Lefo,Iitaka94,Iitaka94b,Iitaka95a,Iitaka95b,Iitaka96,Feit,Kuroda,Raedt1,Raedt2,Kawarabayashi,Natori,Glutsch} has advantage that we can exploit well developed concepts and formalism of quantum theory. It is true especially when we want to deal with quantum systems. Therefore, in this article, we study only the Schroedinger equations.

\subsection{Homogeneous or inhomogeneous?}
\label{subsec:homo}

In this subsection we show that inhomogeneous time-dependent equations are more inefficient than homogeneous ones.  This conclusion is valid not only for the Schroedinger equation (Particle Source Method  \cite{Iitaka96} ) but also for the Newton equation ( Forced Oscillator Method \cite{Williams,Yakubo,Fukamachi,Tanaka,Hukushima}) because both equations are equivalent in the eigenstate representation.
  
Let us define the computational effort of the time-dependent method by the number of timesteps $M=T/\Delta t$. Then the computational effort $M$ is determined by the integration time $T$, because the maximum width of timestep is limited by the {\em sampling theorem} \cite{Recipes} independent of the detail of the method we use. The timestep should be much smaller than the inverse of band width, $\pi/E_B$, to reproduce the correct spectrum since, otherwise, according to (\ref{eq:chi.time.discrete}) we cannot distinguish the eigenvalues    \cite{Kuroda}
\begin{equation}
E_k=E+\frac{2\pi k}{\Delta t} \ \ \ \ (k=1,2,\cdots)
.
\end{equation}

In the following, we evaluate $T$ for homogeneous and inhomogeneous Schroedinger equations to calculate the real-time Green's functions at many frequencies, $\omega_l= l \Delta \omega, \ l=0,\pm 1,\pm 2,\cdots$, within a required relative accuracy $\delta$. It turns out that $T$ of inhomogeneous equations can be much longer than that of homogeneous equations. This conclusion applies to the linear response functions, too.

First let us try to calculate the Green's function by solving the homogeneous equation,
\begin{equation}
\label{eq:schroedinger.inhomogeneous.n}
i \frac{d}{dt}|\phi; t \rangle = H |\phi; t \rangle 
\end{equation}
with the initial condition $|\phi; t=0 \rangle = |j \rangle $. The auxiliary vectors are calculated as 
\begin{eqnarray}
|\tilde{\phi}_l; T \rangle 
%&=& \int_{0}^{T} dt' g^R(t-t') |j \rangle  e^{-i(\omega_l+i\eta)t'}  \\
\label{eq:auxiliary.vector.n.1}
&=& (-i) \int_0^{T} dt' |\phi; t' \rangle e^{+i(\omega_l+i\eta)t'} \\ 
&=& (-i) \int_0^{T} dt' e^{-iHt' } |j \rangle e^{+i(\omega_l+i\eta)t'} \\ 
%&=& (-i) e^{ -iH t }  \int_0^{T} dt' e^{ i(H-\omega_l-i\eta) t'} |j \rangle \\
\label{green.1.1}
&=& \frac{1}{\omega_l+i\eta-H} \left( 1-e^{i(\omega_l+i\eta-H)T} \right)|j \rangle \\
&\approx & \frac{1}{\omega_l+i\eta-H} |j \rangle \\
\label{eq:green.1.2.many}
&=& G(\omega_l+i\eta) |j \rangle 
\end{eqnarray}
where we have neglected the second term of (\ref{green.1.1}) by assuming $T$ is large enough so that $e^{-\eta T} < \delta$.
Therefore we estimate $M$ for the homogeneous equation as
\[
M_1 \approx \frac{ T }{\Delta t}
= \frac{-\log \delta}{\eta \Delta t} 
.
\]

Next let us calculate the Green's function by solving the inhomogeneous Schroedinger equation,
\begin{equation}
\label{eq:schroedinger.inhomogeneous.2}
i \frac{d}{dt}|\phi; t \rangle = H |\phi; t \rangle 
+ |j \rangle \left( \sum_{l=-L}^L e^{-i(\omega_l+i\eta)t} \right) \theta(t)
\end{equation}
with the initial condition $|\phi; t=0 \rangle = 0$.
The solution at large $T$ becomes
\begin{eqnarray}
|\phi; T \rangle
&\approx & \sum_l G(\omega_l+i\eta ) |j \rangle e^{-i(\omega_l +i\eta)T}
\end{eqnarray}
where $T$ satisfies $e^{-\eta T} \ll \delta$.
Then the auxiliary vectors $|\tilde{\phi}_l; T_2 \rangle$ are defined as
\begin{eqnarray}
|\tilde{\phi}_{l'}; T_2 \rangle
&=& \frac{1}{T_2}\int_0^{T_2} dt' | \phi; t \rangle e^{-i(\omega_{l'}+i\eta)t'}
 \nonumber \\
&=&\frac{1}{T_2}\int_0^{T_2} dt' \sum_l  G(\omega_l+i\eta ) |j
\rangle e^{-i(\omega_l-\omega_{l'})t'}  \\
\label{green.m3}
&=& 
 G(\omega_{l'}+i\eta ) |j \rangle 
\nonumber \\
&& + \sum_{ l\ne l'}  G(\omega_l+i\eta ) |j \rangle 
\frac{i\left
( e^{-i(\omega_l-\omega_{l'})T_2}-1\right)}{T_2(\omega_l-\omega_{l'})}
 \\
&\approx& G(\omega_{l'}+i\eta )|j \rangle 
\end{eqnarray}
where we have neglected the second term of (\ref{green.m3}) by assuming that $T_2$ is large enough so that $T_2 \Delta \omega \gg 1/\delta$.
Therefore $M$ becomes
\begin{equation}
\label{eq:many.inhomo.Nprod1}
M_2 \approx 
\frac{1}{\Delta \omega \Delta t \delta} 
\end{equation}
which can be much larger than $M_1$ when $\Delta \omega$ is small.

\section{Non-interacting Electrons}
\label{sec:electron}

In this section, we apply the time-dependent homogeneous Schroedinger equation to calculate efficiently the linear response functions and density of states of non-interacting electron systems, since it is well known that there exist wide and practically important areas in condensed matter physics where non-interacting electron models are useful to predict various physical properties.
Hereafter we assume that the system is
 described by the one-electron Hamiltonian,
\begin{equation}
H=\frac{1}{2} \vec{p}^2 +V(\vec{r})
.
\end{equation}

\subsection{Linear response function}
For non-interacting electrons, 
the linear response function (\ref{eq:chi.eigen}) can be rewritten 
by using one-particle eigenstates as \cite{Imry}
\begin{eqnarray}
\label{eq:chi.eigen.one}
\chi_{BA}(\omega)
&=&
\sum_{E_i \le E_f, E_j>E_f } 
\frac{\langle i| B | j \rangle  \langle j| A | i \rangle}{(\omega+i\eta) -(E_j-E_i )}  \nonumber \\
&&  - \sum_{E_i \le E_f, E_j>E_f } 
\frac{\langle i| A | j \rangle  \langle j | B | i \rangle}{(\omega+i\eta) +(E_j-E_i )} 
\end{eqnarray}
where $E_f$ is the Fermi energy, and $|i\rangle$ and $|j\rangle$ are the occupied and empty one-particle states, respectively.
This formula can be again rewritten in time-dependent representation as
\begin{eqnarray}
\label{eq:chi.time.one}
\lefteqn{
\chi_{BA}(\omega+i\eta) 
} \\
&=& (-i) \int_0^T dt 
\sum_{\begin{array}{c}\scriptstyle E_i \le E_f\\ \scriptstyle E_j>E_f\end{array}} 
 e^{+i(\omega + i\eta)t}
\left\{
\langle i|  e^{+iHt} B e^{-iHt} |j\rangle \langle j| A  | i \rangle 
-c.c.
\right\}
\\
&=& (-i) \int_0^T dt \sum_{E_i,E_j} 
 e^{+i(\omega + i\eta)t} \times \nonumber \\
&& \left\{ \langle i| \theta(E_f-H) e^{+iHt} B e^{-iHt} \theta(H-E_f) |j\rangle \langle j| A  | i \rangle - c.c.\right\}\\
&=& 
\label{eq:chi.numerical.1}
\left\langle \left\langle
\rule{0pt}{24pt}
\int_0^T dt e^{+i(\omega + i\eta)t}  K(t)
\right\rangle \right\rangle
\end{eqnarray}
where the double brackets indicate the statistical average over random vectors $| \Phi \rangle $, and $K(t)$ is the time correlation function defined by
\begin{equation}
\label{eq:correlation}
K(t)= 2 {\rm Im}
\langle \Phi |  \theta(E_f-H)e^{+iHt}B e^{-iHt} \theta(H-E_f)A  | \Phi \rangle 
.
\end{equation}
Equations (\ref{eq:chi.numerical.1}) and (\ref{eq:correlation}) are the main result of this paper.
Note that calculating the trace over the initial states $|i\rangle$ by using random vectors reduces the computational effort by a factor of $N$.
As the result, the computational effort still remains $O(N)$
in spite of the double summation in (\ref{eq:chi.eigen.one}).

In the above equations, we have introduced several numerical techniques.
Firstly, the time-dependent statevectors,
\begin{eqnarray}
\label{eq:timevector}
e^{-iHt} \theta(H-E_f)A  | \Phi \rangle  \nonumber \\
e^{-iHt} \theta(E_f-H) | \Phi \rangle  
\end{eqnarray}
are calculated by the leap frog method \cite{Askar,Lefo,Iitaka94,Iitaka94b,Iitaka96}
\begin{eqnarray}
\label{eq:frog}
|\phi; t + {\Delta t} \rangle &=& -2 i {\Delta t} H |\phi; t \rangle +
|\phi; t - {\Delta t} \rangle 
\end{eqnarray}
where the Hamiltonian matrix is discretized by finite difference \cite{Chelikowsky,Fletcher}
\begin{eqnarray}
%{{\rm \partial }\phi  \over \partial {\mit x}}&=&
%\sum\nolimits\limits_{\mit n\rm =-\mit N_{diff}}^{N_{diff}}
%{C}_{n}^{(1)}\phi \left({{x}+\mit n\Delta x ,{\mit y},{\mit z}}
%\right)
%+\mit O\left({{\Delta x}^{\rm 2\mit N_{diff}}}\right) \\
{{\rm \partial }^{2}\phi  \over \partial {\mit x}^{\rm 2}}&=&
\sum\nolimits\limits_{\mit n\rm =-\mit N_{diff}}^{N_{diff}}
\frac{1}{\Delta x^2}
{C}_{n}^{(2)}\phi \left({{x}+\mit n\Delta x ,{\mit y},{\mit z}}
\right)
+\mit O\left({{\Delta x}^{\rm 2\mit N_{diff}}}\right) 
.
\end{eqnarray}
Due to this discretization, the Hamiltonian matrix becomes sparse and the matrix vector multiplication in (\ref{eq:frog}) can be done with $O(N)$ computational complexity.
We use \( N_{diff}=4 \) formula in this paper.

Secondly, the matrix step function for a normalized hermitian matrix $X$ whose eigenvalues \(X_i \) are in the range $[-1,1]$ is defined in its eigenstate basis
\begin{equation}
\theta(X)= \sum_{X_i} |X_i\rangle \  \theta(X_i) \ \langle X_i |
.
\end{equation}
By using this step function, we can avoid the difficulties in the partial sum in (\ref{eq:chi.time.one}). Operation of this function on an arbitrary vector $|\phi\rangle$ is numerically approximated  by the Chebyshev polynomial expansion  \cite{Recipes,Kosloff,Hoffman,Silver,Roder,Wang,Voter,Goedecker,Sankey},
\begin{equation}
f(X)|\phi\rangle \approx \sum_{k=1}^K c_k T_{k-1}(X)|\phi\rangle
\end{equation}
where each term of the right hand side is calculated by vector recursion formulae
\begin{eqnarray}
T_0(X)|\phi\rangle &=&|\phi\rangle \\
T_1(X)|\phi\rangle &=&X |\phi\rangle \\
T_{n+1}(X)|\phi\rangle &=& 2 X T_n(X)|\phi\rangle -T_{n-1}(X)|\phi\rangle   \ \ n\ge1
.
\end{eqnarray}
To use this matrix function in (\ref{eq:timevector}), we should normalize the Hamiltonian matrix so that \( X=(H-E_f)/E_{norm} \) has eigenvalues in the range \( [-1,1] \).
 
Thirdly, we define random vectors with random phase by
\begin{equation}
\label{eq:random.def}
|\Phi \rangle = \sum_{n=1}^N |n\rangle e^{+i\phi_n} 
\end{equation}
where $|n\rangle $ are basis vectors and 
$-\pi < \phi_n  \le \pi,  (n=1,\cdots,N)$ are uniform random variables that satisfy
\(
\left\langle \left\langle \   e^{-i\phi_{n'}} e^{i\phi_{n}} \  \right\rangle \right\rangle \  = \delta_{n'n}
\).
% The double brackets indicate the statistical average over random variables.
Then we can derive various useful identities such as
\begin{eqnarray}
\label{eq:random.norm}
\langle \Phi | \Phi \rangle &=& 
\sum_n \langle \Phi |n \rangle \langle n | \Phi \rangle
=  \sum_n e^{-i\phi_{n}} e^{i\phi_{n}} =N   \\
\label{eq:random.complete}
\left\langle \left\langle \  | \Phi \rangle \langle \Phi |  \  \right\rangle \right\rangle &=&
\sum_{n'n} |n'\rangle   
\left\langle \left\langle \   e^{-i\phi_{n'}} e^{i\phi_{n}} \  \right\rangle \right\rangle \      \langle n | \nonumber \\ 
&=& \sum_n |n \rangle \langle n | = I \\
\label{eq:trace.monte.appendix}
%\lefteqn{
\left\langle \left\langle \   \langle \Phi | A | \Phi \rangle \  \right\rangle \right\rangle
%}
%\nonumber \\
&=&
\label{eq:random.trace}
\sum_{n, n'} \left\langle \left\langle \   e^{i(\phi_n-\phi_{n'})} \  \right\rangle \right\rangle \  \langle n'|A|n \rangle \nonumber \\
&=&\sum_n \langle n|A|n \rangle  = {\rm tr}\left[ A \right] = \sum_{E_m} \langle E_m|A|E_m \rangle 
\end{eqnarray}
Equation (\ref{eq:random.norm}) shows that each random vector is normalized to $N$, the number of one-particle eigenstates. Equation (\ref{eq:random.complete}) shows that random vectors have normalized completeness. Equation (\ref{eq:random.trace}) shows that the expectation value of an operator by random vectors gives the trace of the operator. We used this identity to calculate the trace over $| i \rangle$ in (\ref{eq:chi.numerical.1}) and (\ref{eq:correlation}).
These random vectors with random {\it phase} are more useful in calculating expectation values than random vectors with random {\it amplitude} since each random vectors are automatically normalized. 

Finally  the formula for numerical calculation of polarizability function $\chi_{\beta \alpha}(\omega)$ with $\alpha,\beta=x,y,z$ becomes
\begin{eqnarray}
\label{eq:chi.dielectric.numerical}
%\lefteqn{
\chi_{\beta \alpha}(\omega) 
%\nonumber \\
&\approx& 
\left\langle \left\langle
\rule{0pt}{24pt}
\int_0^T dt e^{-\eta t} \left( e^{+i \omega t}-\delta_{\beta \alpha} \right)
K(t)
\rule{0pt}{24pt}
\right\rangle \right\rangle \\
\label{eq:chi.correlator.numerical}
K(t)&=&
\frac{-2}{V (\omega+i \eta)^2} {\rm Im}
\langle \Phi |  \theta(E_f-H)e^{+iHt} \times \nonumber \\
&&   p_{\beta} e^{-iHt} \theta(E_{cut}-H) \theta(H-E_f) p_{\alpha}  | \Phi \rangle 
\end{eqnarray}
where $V$ is the volume of the supercell, and the dipole moment operators
\begin{eqnarray}
\label{eq:dipole.operator1}
\langle j |A|i \rangle  &=& \langle j | x_{\alpha} |i \rangle 
 \\
\label{eq:dipole.operator2}
\langle i|B |j \rangle &=&  \frac{-1}{V} \langle i  |x_{\beta}|j \rangle 
,
\end{eqnarray}
are modified to momentum operators by partial integration.
We also inserted a low energy projection operator $\theta(E_{cut}-H)$ into (\ref{eq:chi.correlator.numerical}) to eliminate unphysical high energy components of the random vectors. This filter is much more effective than the quadratic filter used in \cite{Wang}.
In calculating very large systems, we need only few random vectors for statistical averaging, since the fluctuation becomes smaller as the system size $N$ becomes larger \cite{Iitaka96}. 

Figure~\ref{fig:harmonic.eps} shows  the dielectric function 
$\epsilon_{xx}(\omega)=1+ 4\pi \chi_{xx}(\omega)$
of four electrons in  three dimensional harmonic potential
\begin{equation}
\label{eq:harmonic.potential}
V(\vec{r})=\frac{(\omega_0 r)^2}{2}
\end{equation}
calculated with $32^3$ cubic meshes, $\omega_0=0.1$, $\eta=10^{-4}$.
Three random vectors are used.
The analytical result \cite{Jackson} 
\begin{equation}
\epsilon_{xx}(\omega)=1+ \frac{4\pi N_e}{V} 
\frac{1}{\omega_0^2-\omega^2-i\omega \eta}
\end{equation}
is also shown for comparison, 
where $N_e$ is the number of electrons in the supercell of volume $V$.
The deviation from the exact result near $\omega=0$ is due to finite $\eta$.
The result shows that our method works very well for $\omega \gg \eta$.

Figure~\ref{fig:silicon.crystal.eps}  shows  the dielectric function 
with energy resolution $\eta=0.05 (eV)$
 of silicon crystal consisting of $2^{15}$ Si atoms in a cubic supercell of $16^3$ unit cells. Each unit cell is divided into $8^3$ cubic meshes. One random vector is used. We used the empirical local pseudopotential in reference \cite{Zunger}.
The result agrees with experimental results and other
theoretical calculations \cite{Cohen,Noguez}.

In some cases we may want to ask which part of the real space 
the electrons contributing to the linear response function come from.
We can answer to this question by calculating the linear response function by restricting the range of the trace in (\ref{eq:chi.dielectric.numerical}) within a real space domain $D$. This can be done by replacing $|\Phi\rangle$ 
by $|\Phi'\rangle=P_D|\Phi\rangle$ where $P_D=\sum_{n \in D} |n\rangle \langle n|$ is the real space projection operator onto $D$.

\subsection{Density of states}

The density of states of the system can be calculated as  \cite{Raedt2}
\begin{eqnarray}
\label{eq:dos}
\rho(\omega) &=& \frac{-1}{\pi} \sum_n {\rm Im \ } G_{nn}(\omega + i\eta) 
= \frac{-1}{\pi} {\rm Im \ } \left( {\rm tr} \left[ G(\omega + i\eta) \right] \right)
\end{eqnarray}
by combining (\ref{eq:green.1.2.many}) and (\ref{eq:trace.monte.appendix}).

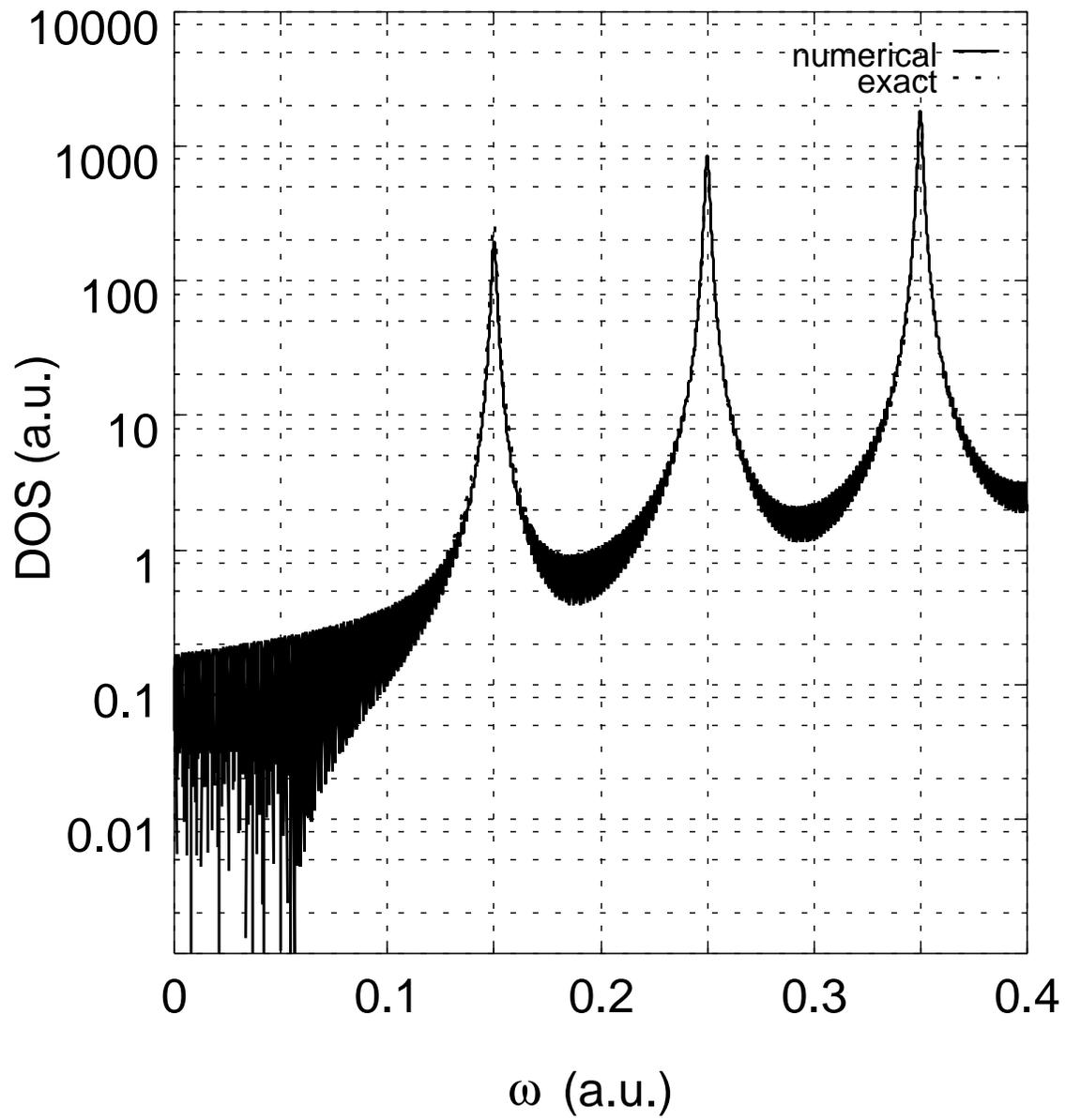
\begin{figure}
\caption{
Density of states of 3D harmonic oscillator calculated
with $32^3$ cubic meshes, $\omega_0=0.1$, and  $\eta=10^{-4}$,
and analytical result.
}
\label{fig:harmonic.dos}
\end{figure}

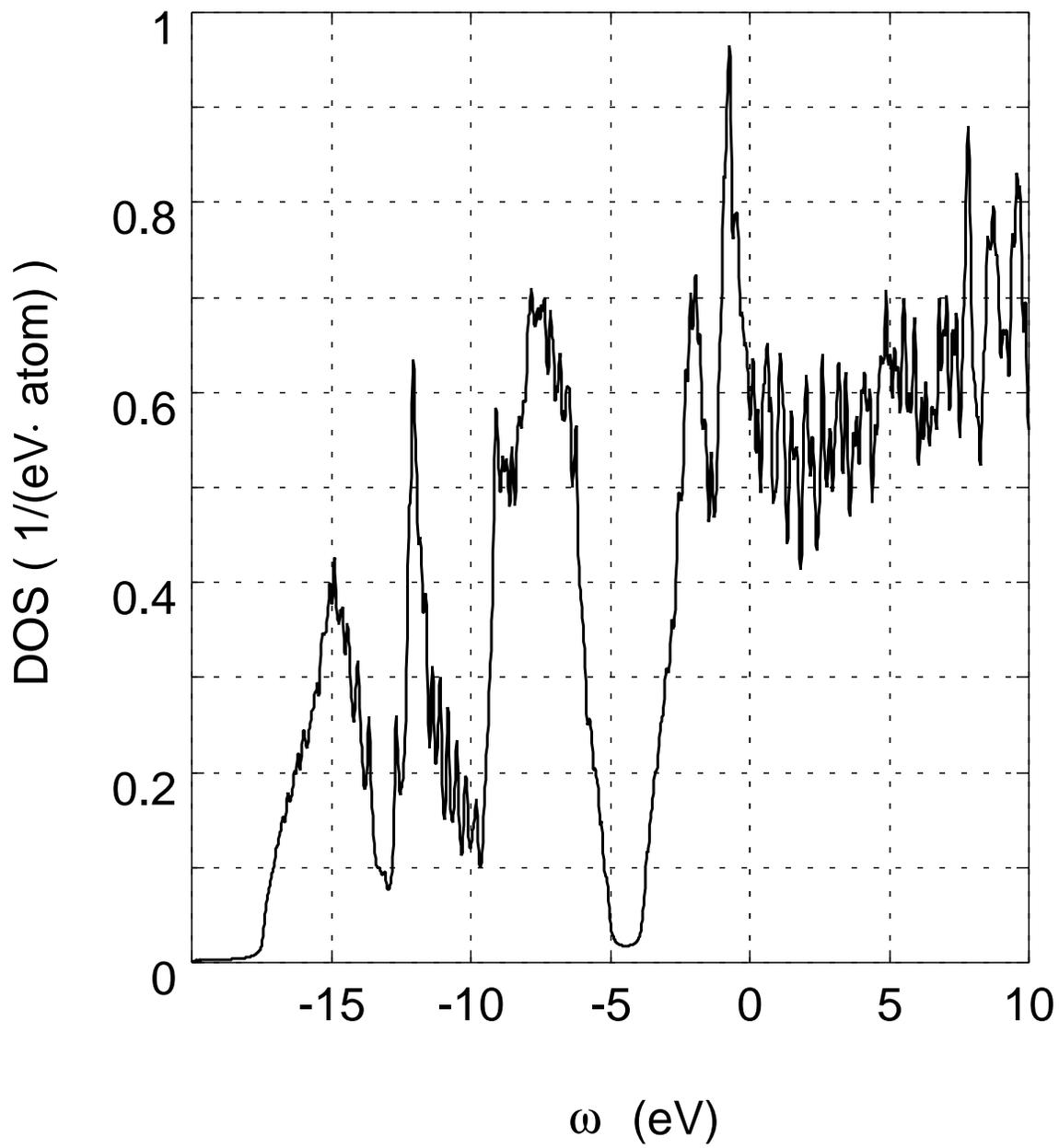
\begin{figure}
\caption{
Density of states
 of 
 silicon crystal consisting of $2^{15}$ Si atoms in $16^3$ unit cells. Each unit cell is divided into $8^3$ cubic meshes.
The energy resolution is $\eta=0.05 (eV)$.
}
\label{fig:crystal.dos}
\end{figure}

Figure~\ref{fig:harmonic.dos} shows  the numerical and analytical results of  the density of states in 3D harmonic potential 
with $32^3$ cubic meshes, $\omega_0=0.1$, and  $\eta=10^{-3}$.
Three random vectors are used.
Figure~\ref{fig:crystal.dos} shows  the density of states
 of silicon crystal consisting of $2^{15}$ Si atoms in a cubic supercell of $16^3$ unit cells. Each unit cell is divided into $8^3$ cubic meshes.
The energy resolution is $\eta=0.05 (eV)$. Three random vectors are used.

%\subsection{Local density of states}
We can also calculate the {\it local} density of states integrated in a given domain
$D$ by using the real space projection operator $P_D$ to restrict the summation in (\ref{eq:random.def}) within $D$,
\begin{eqnarray}
\label{eq:localdos}
\rho_D(\omega) &=& \frac{-1}{\pi} \sum_{n \in D} {\rm Im \ } G_{nn}(\omega + i\eta) 
= \frac{-1}{\pi} {\rm Im \ } \left( {\rm tr} \left[ P_D G(\omega + i\eta) \right] \right)
.
\end{eqnarray}

Photonic band structures in two-dimensional periodic structure of dielectric material \cite{Plihal,Baba,Hirayama} can also be calculated by using (\ref{eq:dos}) or (\ref{eq:localdos}) since
the Maxwell equations of this system are reduced to the Schroedinger equation with position dependent mass, i.e. ,
\begin{eqnarray}
\label{eq:photonic.hamiltonian}
{\cal H} H_z(x,y)
&=&\frac{\omega^2}{c^2} H_z(x,y) = E H_z(x,y) \\
{\cal H} &=&
\frac{\partial}{\partial x}\frac{-1}{\epsilon(x,y)} \frac{\partial }{\partial x}
+\frac{\partial}{\partial y}\frac{-1}{\epsilon(x,y)} \frac{\partial }{\partial y}
\end{eqnarray}
for $H$-mode where $H_z$ is the $z$-component of the magnetic field, and
\begin{eqnarray}
\label{eq:photonic.hamiltonian.emode}
{\cal H} E_z(x,y)
&=&\frac{\omega^2}{c^2} E_z(x,y) = E E_z(x,y) \\
{\cal H} &=&
\frac{-1}{\epsilon(x,y)} \left\{
\frac{\partial^2 }{\partial x^2}
+\frac{\partial^2}{\partial y^2}
\right\}
\end{eqnarray}
for $E$-mode where $E_z$ is the $z$-component of the electric field.

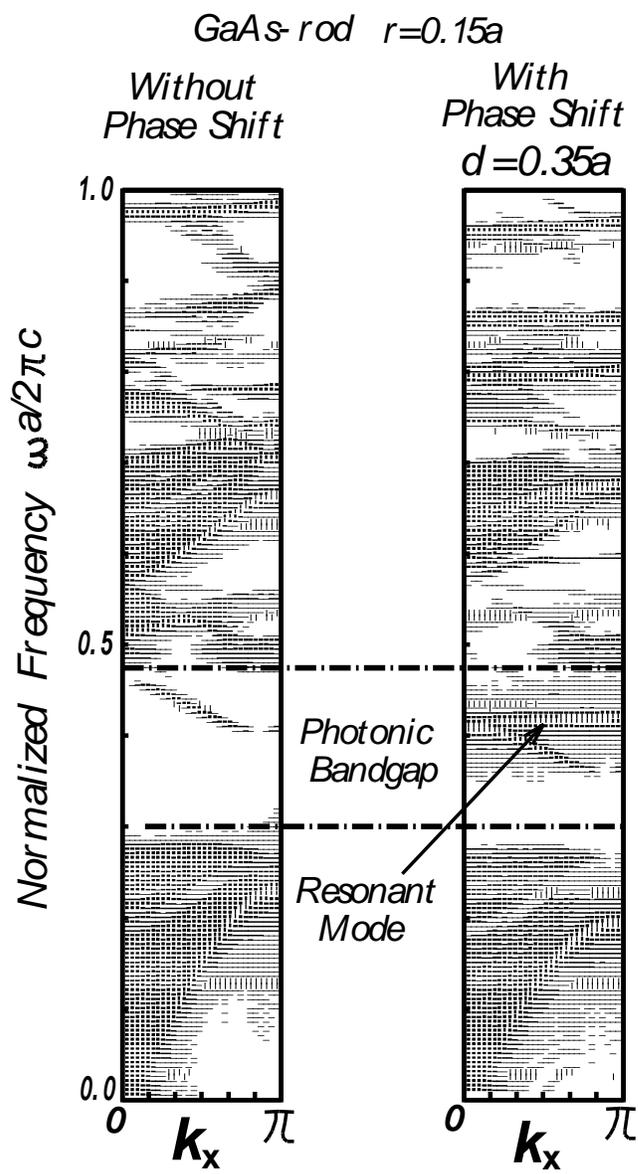
\begin{figure}
\caption{
The calculated density of states as a function of frequency
and wave number
}
\label{fig:6}
\end{figure}

Figure~\ref{fig:5} shows a typical structure of two-dimensional
photonic crystal cavities used in our calculation, and Fig.~\ref{fig:6}
shows the calculated density of states as a function of frequency
and wave number.

\section{Summary}
In this article we proposed a new numerical method suitable for calculating the linear response functions (and the density of states) of non-interacting electrons, in which the sum over the initial one-particle states are efficiently calculated by using random vectors. 
The advantage of this method compared to the Chebyshev polynomial method by Wang to calculate optical absorption of non-interacting electrons \cite{Wang} is that our method can calculate not only the imaginary part but also the real part of the linear response functions at the same time, and  that it can calculate them without any input-output of statevectors on external storage. As the result, our method can calculate much larger systems than Wang's method.  The Chebyshev polynomial method of degree $M$ should store $O(M)$ statevectors of size $O(N)$ on external storage to make the table of $O(M^2)$ generalized Chebyshev moments $\Lambda_{m,m'}$ and may take very long I/O time. 

The application of this method to photonic band structures, silicon nanocrystalites, and periodic structures of chaotic systems will be presented elsewhere \cite{LDSD1,LDSD2,LDSD3}.

\section*{Acknowledgments}
One of the authors (T.I) wishes to thank Prof. Masuo Suzuki and the referee of this article for their encouragement.
The calculations were performed on Fujitsu VPP500 at RIKEN and ISSP.

\clearpage

\begin{table}
\[
\begin{array}{c|c|c|c|c}
\multicolumn{5}{c}{\mbox{Diagonalization Method}} \\
\hline
\multicolumn{5}{c}{
\begin{array}{lll}
\displaystyle
\chi_{BA}(\omega+i\eta)
&=&
\displaystyle
\sum_{m=1}^N 
\frac{\langle E_g| B | E_m \rangle  \langle E_m| A | E_g \rangle}{(\omega+i\eta) -(E_m-E_g )}  
 - \sum_{m=1}^N 
\frac{\langle E_g| A | E_m \rangle  \langle E_m| B | E_g \rangle}{(\omega+i\eta) +(E_m-E_g )} 
\end{array}
} \\
\hline
  & \multicolumn{2}{c|}{\mbox{Dense Matrix}}& \multicolumn{2}{c}{\mbox{Sparse Matrix}}\\
\cline{2-5}
\raisebox{6pt}[0pt]{\makebox[60mm]{Calculation}}& \mbox{Computation} & \mbox{Memory}& \mbox{Computation}& \mbox{Memory} \\
\hline
E_m, |E_m\rangle           & N^3 & N^2 & N^3 & N^2 \\
\langle E_g| A |E_n\rangle & N^2 & N   & N   & N   \\
\displaystyle
\sum_{m} 
                         & N   & N   & N   & N   \\
\hline
\end{array}
\]

\[
\begin{array}{c|c|c|c|c}
\multicolumn{5}{c}{\mbox{Time-dependent Method}} \\
\hline
\multicolumn{5}{c}{
\begin{array}{lll}
\displaystyle
\chi_{BA}(\omega+i\eta) &=&
\displaystyle
2  \sum_{m=0}^M \Delta t e^{+i(\omega+E_g +i\eta)m \Delta t}
{\rm Im} \langle E_g| B e^{-iH m\Delta t} A | E_g \rangle 
\end{array}
} \\
\hline
  & \multicolumn{2}{c|}{\mbox{Dense Matrix}}& \multicolumn{2}{c}{\mbox{Sparse Matrix}}\\
\cline{2-5}
\raisebox{6pt}[0pt]{\makebox[60mm]{Calculation}}& \mbox{Computation}& \mbox{Memory}& \mbox{Computation}& \mbox{Memory} \\
\hline
e^{-iHm\Delta t} B | E_g \rangle               & M N^2 & N^2 & M N & N \\
\langle E_g| A e^{-iHm\Delta t} B  |E_g\rangle & N^2   & N   & N   & N   \\
\displaystyle
\sum_{m=0}^M 
                         & M   & 1   & M   & 1   \\
\hline
\end{array}
\]

\caption{Comparison of diagonalization method and time-dependent method}
\label{tbl:compare}
\end{table}

\clearpage

\begin{table}
\begin{tabular}{l|l|l|l}
\hline
           & equation           & homogeneous    & inhomogeneous \\
\hline
\hline
classical mechanics   & Newton        &  harmonic oscillator   & forced oscillator \\
           &                      &                     & \cite{Williams,Yakubo,Fukamachi,Tanaka,Hukushima} \\
\hline
quantum mechanics   & Schroedinger  & TDSE                & particle source \\
           &                       &   \cite{Askar,Lefo,Iitaka94,Iitaka94b,Iitaka95a,Iitaka95b}  & \cite{Iitaka96} \\
           &                       &   \cite{Feit,Raedt1,Raedt2,Glutsch}  &  \\
\hline
\end{tabular}

\caption{Comparison of time-dependent equations}
\label{tbl:equations}
\end{table}

\clearpage

\clearpage

\begin{figure}
\caption{
$\epsilon_{xx}(\omega)$  of four electrons in 
a three dimensional harmonic oscillator
calculated with $32^3$ cubic meshes, $\omega_0=0.1$, $\eta=10^{-4}$;
(a) real part, (b) imaginary part.
}
\label{fig:harmonic.eps}
\end{figure}

\begin{figure}
\caption{
$\epsilon_{xx}(\omega)$ of silicon crystal consisting of $2^{15}$ Si atoms in a cubic supercell of $16^3$ unit cells. Each unit cell is divided into $8^3$ cubic meshes. The energy resolution is $\eta= 0.05 (eV)$. We used the empirical local pseudopotential in reference \protect\cite{Wang}.
(a) real part, (b) imaginary part.
}
\label{fig:silicon.crystal.eps}
\end{figure}

%\begin{figure}
%\caption{Density of states of amorphous silicon}
%\label{fig:silicon.amorphous.eps}
%\end{figure}

%\begin{figure}
%\caption{Errors in density of states}
%\label{fig:silicon.error.eps}
%\end{figure}

\begin{figure}
\caption{
A typical structure of two-dimensional
photonic crystal cavities used in our calculation
}
\label{fig:5}
\end{figure}

\end{document}